\documentclass[12pt]{article}
\usepackage{latexsym}
\usepackage[dvips]{graphicx}

\def\ba{\begin{array}}
\def\ea{\end{array}}
\def\be{\begin{equation}\begin{array}{l}}
\def\ee{\end{array}\end{equation}}
\def\c{\cite}

\begin{document}
\vskip 2cm
\noindent
\begin{center}
 {\large\bf   A Note on Symplectic, Multisymplectic  \\\vskip 3mm
Scheme in
 Finite Element Method}

\end{center}
\vspace*{1cm}

\def\thefootnote{\fnsymbol{footnote}}
\begin{center}{\sc Han-Ying GUO${}^1$}\footnote{email: hyguo@itp.ac.cn},
{\sc Xiao-mei JI ${}^{1,2}$}\footnote{email: jixm@iu-math.math.indiana.edu }
{\sc Yu-Qi LI${}^{1}$}\footnote{email: qylee@itp.ac.cn}
 and {\sc Ke WU${}^1$}\footnote{email: wuke@itp.ac.cn} 
\end{center}
\begin{center}
{\it ${}^1$ Institute of Theoretical Physics, Academia Sinica, P.O.Box 2735,\\ Beijing 
100080, China \par
${}^2$ Department of Mathematics, Indiana University\\ Bloomington, IN 47405, U.S.A.}
\end{center}

\vskip 2mm
\vfill

\noindent
\centerline {\sc Abstract}
\vskip 3mm

$\quad $ We
find that with uniform mesh, the numerical schemes derived
from finite element method can keep a preserved symplectic structure
in one-dimensional case and a preserved multisymplectic structure in
two-dimentional case in certain discrete version respectively.
These results are in fact the intrinsic reason
that the numerical experiments indicate
that such finite element algorithms are accurate in practice.

\vskip .8cm
 
\vskip .8cm

\noindent

\vskip .4cm
\newpage

 $\quad $   As both the  finite elemente method\c{fini} and
 symplectic scheme \c{symp} as well as multisymplectic scheme \c{msymp} are
 powerful tools to solve differential equations numericaly, it is interesting to 
explore 
 if there  is some relation between them.  We will give a partial answer
 to this question by a simple example.

 In order to show  the symplectic or multisymplectic structures in the
 scheme derived from finite elemente method , we consider the boundary
 value problem of the
 semi-linear elliptic equation in one-dimensional and two-dimensional spaces:
\begin{equation}
\triangle u=f(u) \quad in \quad \Omega,
\quad u|_{\partial{\Omega}}=0 \quad on \quad \partial\Omega.
\end{equation}
where $\Omega $ is a bounded domain in ${I\!\!R}^n$, $n=1,2$ and $f(u)$
is nonlinear and sufficiently smooth enough function.
The weak formulation of the boundary value problem of the equation is
to find $u:\Omega\rightarrow H^1_0(\Omega)$ such that
\begin{equation}\label{3}
\int _{\Omega}\nabla u\cdot\nabla v dx=
-\int _{\Omega}f\cdot v dx \hskip
0.5cm \forall v\in H^1_0(\Omega),
\end{equation}
Let
$$
a(u,v)=\int _{\Omega}\nabla u\cdot\nabla v dx,\quad
(f,v)=-\int _{\Omega}f\cdot v dx,
$$
then (\ref{3}) becomes,
\begin{eqnarray}
a(u,v)= (f,v).
\end{eqnarray}

It is important to note that the equation is in
fact an ODE with Lagrangian on
${I\!\!R}^1$ or Lagrangian PDE in  ${I\!\!R}^2$ respectively. The details
of the symplectic and multisymplectic structures in the
Lagrangian formalism can be found in \c{symp} \c{msymp} \c{glw1} \c{glw2}
\c{vese} \c{mars} \c{qin}.

 In one-dimensional case  we first discrete ${I\!\!R}^1$ with regular lattice
${I\!\!L}^1$ with equal spatial step $h$
$$
{I\!\!L}^1 =\{\cdots, x_{i-1}, x_i, x_{i+1}, \cdots\}
$$
 Let $\Omega$ be  a segment in ${I\!\!R}^1$ and $\varphi _i$ linear shape function such that $\varphi _i(x_j)=\delta _{ij}$. %For given
As usual, $u_i=u(x_i)$.

From the finite element method, we get the scheme
\begin{eqnarray}
\frac{u_{i+1}-2u_i+u_{i-1}}{h}=
\int_{i-1}^{i+1}f(\sum_{i-1}^{i+1}u_k\varphi_k)\varphi_idx.
\end{eqnarray}
The right side can be rewritten as
\begin{eqnarray}
I_i:=\int_{i-1}^{i}f(u_{i-1}\varphi_{i-1}+u_{i}\varphi_{i})\varphi_idx+
\int_{i}^{i+1}f(u_{i}\varphi_{i}+u_{i+1}\varphi_{i+1})\varphi_idx.
\end{eqnarray}

It should be noted that the variables $u_i$'s in the scheme 
can be released from the solution
space to the function space by means of relevant discrete Euler-Lagrange (DEL) 
cohomologically
equivalent relation \c{glw1}.
Therefore, as long as working with the DEL cohomology class associated with the DEL 
euation (times by certain 1-form) the $u_{k}$'s can be regarded as in
the function space in general rather than in the solution space. 

Introducing the DEL 1-forms
\begin{eqnarray}
E_{D i} :=\{{u_{i+1}-2u_i+u_{i-1}}-hI_i\}du_i,
\end{eqnarray}
such that the null DEL 1-form gives rise to the equation in the finite
element
method. The DEL condition reads
\begin{eqnarray}
dE_{D i} =0. 
\end{eqnarray}
Namely, the DEL 1-forms are closed. 
It is straightforward to see that
from the DEL condition it follows
\begin{eqnarray}
&&du_{i+1}\wedge du_i+du_{i-1}\wedge du_i
\nonumber\\
&=&h(\int_{i-1}^{i}f^{\prime}(u_{i-1}\varphi_{i-1}+u_{i}\varphi_{i})
\varphi_{i-1}\varphi_idx)du_{i-1}\wedge du_i\nonumber\\
&\quad +&h(\int_{i}^{i+1}f^{\prime}(u_{i}\varphi_{i}+u_{i+1}\varphi_{i+1})
\varphi_{i+1}\varphi_idx)du_{i+1}\wedge du_i ,
        \end{eqnarray}
i.e.,
\begin{eqnarray}
\omega^{(i+1)}=\omega^{(i)}.
 \end{eqnarray}
where 
\begin{eqnarray}
\omega^{(i+1)}
=(1-h\int_{i}^{i+1}f^{\prime}(u_{i}\varphi_{i}+u_{i+1}\varphi_{i+1})
\varphi_{i+1}\varphi_idx)du_{i+1}\wedge du_i.
\end{eqnarray}
It can be checked that this 2-form is closed w.r.t. $d$
on the function space and its coefficients are non-degenerate,
so that it is a symplectic structure for the scheme derived
from finite element method in one-dimensional and it is preserved.

   In the two-dimensional case, the semi-linear equation becomes
\begin{equation}
 u_{x_1x_1}+u_{x_2x_2}=f(u) \quad in \quad \Omega,
\quad u|_{\partial{\Omega}}=0 \quad on \quad \partial\Omega.
\end{equation}
where $u_{x_i}$, $i=1,2$, denote the partial derivative of
$u$ w.r.t. coordinate $\{x_i\}$ in ${I\!\!R}^2$.
 Without loss of generality, we assume $\Omega$ is a square
domain, and the mesh
is uniform, that is, the plane ${I\!\!R}^2$ is divided into squares
$\{(x_1, x_2);i_1h\leq x_1\leq (i_1+1)h, i_2h\leq x_2\leq (i_2+1)h\},
i_1,i_2=0,\pm 1,\pm 2
,\cdots$, and each square is further divided into two triangles by a
straight line $x_2=x_1+ih$, $i$ integer.
Take the node $x_{i,j}$ and $\Omega _{i,j}$ shown as figure 1. The elements
are divided  into two categories: the first  category is shown as figure 2
and the second is shown as figure 3.

\begin{figure}
\begin{center}
\begin{minipage}{5cm}
\includegraphics[width=5cm]
{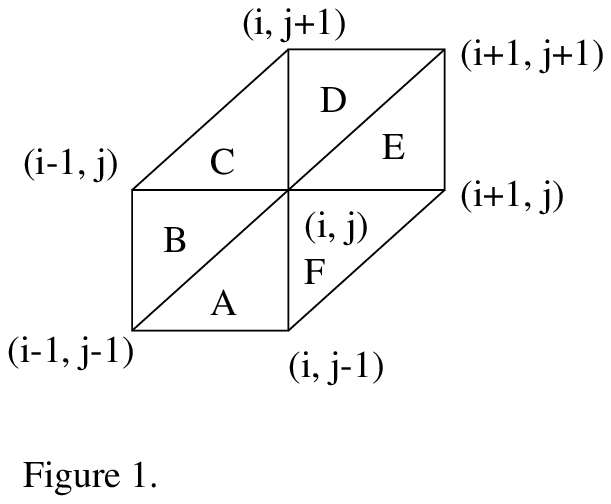}
\caption{}

\end{minipage}
\hskip3mm
\begin{minipage}{3.5cm}
\includegraphics[width=3.5cm]
{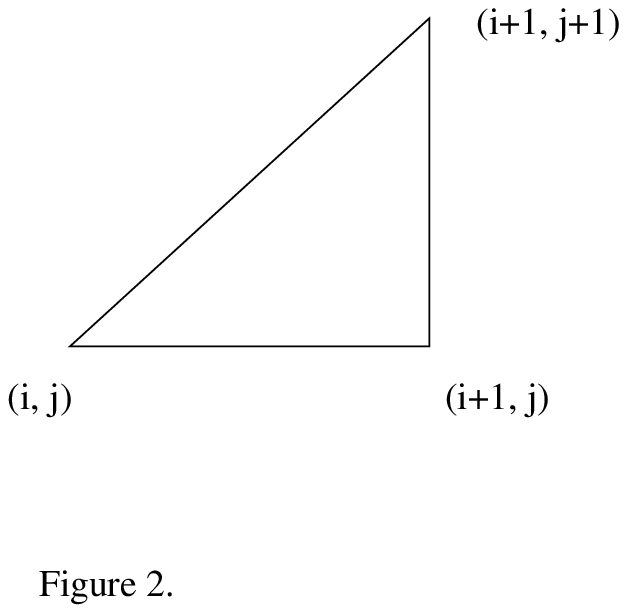}
\caption{}

\end{minipage}

\begin{minipage}{3.5cm}
\includegraphics[width=3.5cm]
{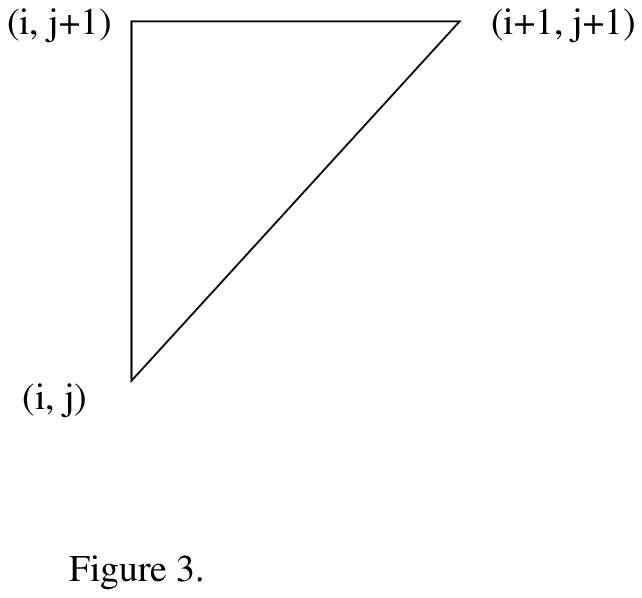}
\caption{}

\end{minipage}
        
\end{center}
\end{figure}

The discrete scheme for this equation form finite element method is
\begin{eqnarray}
\label{fin0}
        u_{i,j+1}+u_{i+1,j}+u_{i,j-1}+u_{i-1,j}-4u_{i,j}\nonumber\\
=\int_{\Omega_{i,j}}
f(\sum u_{k,l}\varphi_{k,l})\varphi_{i,j}dx,
\end{eqnarray}
where the shape function $\varphi_{i,j}$ is linear continuous function
defined as
$$
\varphi_{i,j}(x_{k,l})=\delta_{i.k}\delta_{j,l}.
$$
The right side of (\ref{fin0}) is
\begin{eqnarray}
I_{\Omega_{i,j}}&:=&\int_{\Omega_{i,j}} f(\sum u_{k,l}\varphi_{k,l})
\varphi_{i,j}dx\nonumber\\
&=&(\int_A+\int_B+\int_C+\int_D+\int_E+\int_F)f(\sum u_{k,l}\varphi_{k,l})
\varphi_{i,j}dx,
\end{eqnarray}
where
\begin{eqnarray}
I_{\Omega_{i,j}A}&:=&\int_Af(\sum u_{k,l}\varphi_{k,l})
\varphi_{i,j}dx\nonumber\\
&=&\int_Af(u_{i-1,j-1}\varphi_{i-1,j-1}+
u_{i,j-1}\varphi_{i,j-1}+u_{i,j}\varphi_{i,j})\varphi_{i,j}dx.\nonumber
\end{eqnarray}
\begin{eqnarray}
I_{\Omega_{i,j}B}&:=&\int_Bf(\sum u_{k,l}\varphi_{k,l})
\varphi_{i,j}dx\nonumber\\
&=&\int_Bf(u_{i-1,j-1}\varphi_{i-1,j-1}+
u_{i-1,j}\varphi_{i-1,j}+u_{i,j}\varphi_{i,j})\varphi_{i,j}dx\nonumber
\end{eqnarray}
\begin{eqnarray}
I_{\Omega_{i,j}C}&:=&\int_Cf(\sum u_{k,l}\varphi_{k,l})
\varphi_{i,j}dx\nonumber\\
&=&\int_Cf(u_{i-1,j}\varphi_{i-1,j}+
u_{i,j}\varphi_{i,j}+u_{i,j+1}\varphi_{i,j+1})\varphi_{i,j}dx\nonumber
\end{eqnarray}
\begin{eqnarray}
I_{\Omega_{i,j}D}&:=&\int_Df(\sum u_{k,l}\varphi_{k,l})
\varphi_{i,j}dx\nonumber\\
&=&\int_Df(u_{i+1,j+1}\varphi_{i+1,j+1}+
u_{i,j}\varphi_{i,j}+u_{i,j+1}\varphi_{i,j+1})\varphi_{i,j}dx\nonumber
\end{eqnarray}
\begin{eqnarray}
I_{\Omega_{i,j}E}&:=&\int_Ef(\sum u_{k,l}\varphi_{k,l})
\varphi_{i,j}dx\nonumber\\
&=&\int_Ef(u_{i+1,j+1}\varphi_{i+1,j+1}+
u_{i,j}\varphi_{i,j}+u_{i+1,j}\varphi_{i+1,j})\varphi_{i,j}dx\nonumber
\end{eqnarray}
\begin{eqnarray}
I_{\Omega_{i,j}F}&:=&\int_Ff(\sum u_{k,l}\varphi_{k,l})
\varphi_{i,j}dx\nonumber\\
&=&\int_Ff(u_{i+1,j}\varphi_{i+1,j}+
u_{i,j}\varphi_{i,j}+u_{i,j-1}\varphi_{i,j-1})\varphi_{i,j}dx.\nonumber
\end{eqnarray}
where the sub-index $A,B,C,D,E,F$ indicate the all elements neighboring
$x_{i,j}$ in Figure 1.

Similar to the one-dimensional case, introducing the DEL 1-forms
\begin{equation}
E_{D \Omega_{i,j}}:=\{ u_{i,j+1}+u_{i+1,j}+u_{i,j-1}+u_{i-1,j}-4u_{i,j}
- I_{\Omega_{i,j}}\}du_{i,j}.
\end{equation}
The DEL condition now reads
\begin{equation}
dE_{D \Omega_{i,j}}=0, %d\a _{D \Omega_{i,j}},
\end{equation}
i.e.
 \begin{eqnarray}
\label{fin1}
&&du_{i,j+1}\wedge du_{i,j}+du_{i+1,j}\wedge du_{i,j}
+du_{i,j-1}\wedge du_{i,j}+du_{i-1,j}\wedge du_{i,j}\nonumber\\
&&= S_{A\{i,j\}}+ S_{B\{i,j\}} +S_{C\{i,j\}}+ S_{D\{i,j\}}+ S_{E\{i,j\}}+ S_{F\{i,j\}}
\end{eqnarray}
where,
\begin{eqnarray}
S_{A\{i,j\}}
&=&(\int_Df^\prime(u_{i+1,j+1}\varphi_{i+1,j+1}+
u_{i,j}\varphi_{i,j}+u_{i,j+1}\varphi_{i,j+1})
\varphi_{i+1,j+1}\varphi_{i,j}dx+\nonumber\\
&&\int_Ef^\prime(u_{i+1,j+1}\varphi_{i+1,j+1}+
u_{i,j}\varphi_{i,j}+u_{i+1,j}\varphi_{i+1,j})
\varphi_{i+1,j+1}\varphi_{i,j}dx  
)\cdot du_{i+1,j+1}\wedge du_{i,j},\nonumber\\
S_{B\{i,j\}}
&=&(\int_Df^\prime(u_{i+1,j+1}\varphi_{i+1,j+1}+
u_{i,j}\varphi_{i,j}+u_{i,j+1}\varphi_{i,j+1})
\varphi_{i,j+1}\varphi_{i,j}dx+\nonumber\\
&&\int_Cf^\prime(u_{i-1,j}\varphi_{i-1,j}+
u_{i,j}\varphi_{i,j}+u_{i,j+1}\varphi_{i,j+1})
\varphi_{i,j+1}\varphi_{i,j}dx
)\cdot du_{i,j+1}\wedge du_{i,j},\nonumber\\  
S_{C\{i,j\}}
&=&(\int_Ef^\prime(u_{i+1,j+1}\varphi_{i+1,j+1}+
u_{i,j}\varphi_{i,j}+u_{i+1,j}\varphi_{i+1,j})
\varphi_{i+1,j}\varphi_{i,j}dx+\nonumber\\
&&\int_Ff^\prime(u_{i+1,j}\varphi_{i+1,j}+
u_{i,j}\varphi_{i,j}+u_{i,j-1}\varphi_{i,j-1})
\varphi_{i+1,j}\varphi_{i,j}dx
)\cdot du_{i+1,j}\wedge du_{i,j},\nonumber\\ 
S_{D\{i,j\}}
&=&(\int_Ff^\prime(u_{i+1,j}\varphi_{i+1,j}+
u_{i,j}\varphi_{i,j}+u_{i,j-1}\varphi_{i,j-1})
\varphi_{i,j-1}\varphi_{i,j}dx+\nonumber\\
&&\int_Af^\prime(u_{i-1,j-1}\varphi_{i-1,j-1}+
u_{i,j-1}\varphi_{i,j-1}+u_{i,j}\varphi_{i,j})
\varphi_{i,j-1}\varphi_{i,j}dx
)\cdot du_{i,j-1}\wedge du_{i,j},\nonumber\\     
S_{E\{i,j\}}
&=&(\int_Af^\prime(u_{i-1,j-1}\varphi_{i-1,j-1}+
u_{i,j-1}\varphi_{i,j-1}+u_{i,j}\varphi_{i,j})
\varphi_{i-1,j-1}\varphi_{i,j}dx+\nonumber\\
&&\int_Bf^\prime(u_{i-1,j-1}\varphi_{i-1,j-1}+
u_{i-1,j}\varphi_{i-1,j}+u_{i,j}\varphi_{i,j})
\varphi_{i-1,j-1}\varphi_{i,j}dx)\cdot du_{i-1,j-1}\wedge du_{i,j},\nonumber
\\
S_{F\{i,j\}}
&=&(\int_Bf^\prime(u_{i-1,j-1}\varphi_{i-1,j-1}+
u_{i-1,j}\varphi_{i-1,j}+u_{i,j}\varphi_{i,j})
\varphi_{i-1,j}\varphi_{i,j}dx+\nonumber\\
&&\int_Cf^\prime(u_{i-1,j}\varphi_{i-1,j}+
u_{i,j}\varphi_{i,j}+u_{i,j+1}\varphi_{i,j+1})
\varphi_{i-1,j}\varphi_{i,j}dx
)\cdot du_{i-1,j}\wedge du_{i,j}.\nonumber
\end{eqnarray}

Let us introduce two shift operators as
\begin{eqnarray}
E_1f(u_{i,j})=f(u_{i+1,j}), \nonumber\\
E_2f(u_{i,j})=f(u_{i,j+1}). \nonumber
\end{eqnarray}
Then  the following relations can be found
\begin{eqnarray}
S_{A\{i,j\}}&=&-E_1E_2S_{E\{i,j\}}, \nonumber\\
S_{B\{i,j\}}&=&-E_2S_{D\{i,j\}}, \nonumber\\
S_{C\{i,j\}}&=&-E_1S_{F\{i,j\}}, \nonumber\\
du_{i,j+1}\wedge du_{i,j}&=&-E_2(du_{i,j-1}\wedge du_{i,j}),\nonumber\\
du_{i+1,j}\wedge du_{i,j}&=&-E_1(du_{i-1,j}\wedge du_{i,j}).\nonumber
\end{eqnarray}
Then it is easy to check that (\ref{fin1}) becomes
\begin{eqnarray}
\label{cons}
D_1 \omega_{D{i,j}} +D_2 \tau_{D{i,j}} =0. %\nonumber
\end{eqnarray}
where
\begin{eqnarray}
\omega_{D{i,j}}=du_{i-1,j}\wedge du_{i,j}-E_2 S_{E\{i,j\}}-S_{F\{i,j\}},\\
\tau_{D{i,j}}=du_{i,j-1}\wedge
du_{i,j}-S_{B\{i,j\}}-S_{E\{i,j\}}.
\end{eqnarray}
It is straightforward to show that $\omega_D$ and $\tau_D$ are
two symplectic 2-forms. Namely, they are closed w.r.t. $d$ on the
function space and non-degenerate. Then the
equation (\ref{cons}) is in fact  the multisymplectic
conservation law. 
Here $D_1$ and $D_2$ are given by
\begin{equation}
D_1=E_1-1,\quad D_2=E_2-1. \quad
\end{equation}

  In this letter we have  explored some very interesting relations
 between symplectic and
 multisymplectic algorithms and the simple 
finite element method for the  boundary value problem of the
 semi-linear elliptic equation in one-dimensional and two-dimensional spaces.
The details you could found in \cite{gjlw}. 
Although what we have found are certain simple boundary value problem of
the semi-linear  elliptic equation in lower dimensions and also quite simple 
triangulization in the
finite element method, the results still indicate that there should be very 
deep connections 
between the symplectic or multisymplectis algorithms and the finite
 element method.  

 There are lots of  relevant problems should to be studied 
and some of them are under investigation.

\end{document}